\documentclass[12pt]{article}
\usepackage{setspace}
\usepackage{amsmath,amssymb}
\usepackage{graphicx}
\usepackage{subfig}
\begin{document}

\newcommand{\beq}{\begin{equation}}
\newcommand{\eeq}[1]{\label{#1}\end{equation}}
\newcommand{\bea}{\begin{eqnarray}}
\newcommand{\eea}[1]{\label{#1}\end{eqnarray}}
\def\draftnote#1{{\color{red} #1}}
\def\bldraft#1{{\color{blue} #1}}

\numberwithin{equation}{section}

\begin{titlepage}
\begin{center}

\vskip 4 cm

{\Large \bf More on Long String Dynamics in Gravity on AdS$_3$: Spinning Strings and Rotating BTZ }

\vskip 1 cm

Jihun Kim$^a$  and Massimo Porrati$^{a,b}$

\vskip .75 cm

{$^a$ \em Center for Cosmology and Particle Physics, \\ Department of Physics, New York University, \\4 Washington Place, New York, NY 10003, USA}

\vskip .75 cm

{$^b$ \em School of Natural Sciences, Institute for Advanced Study\\ Princeton NJ 
USA 08540}\footnote{Member until May 2015, on sabbatical leave from NYU.}
\end{center}

\vskip 1.25 cm

\begin{abstract}
\noindent
In this paper we study the classical dynamics of long strings in AdS$_3$, generalizing  our previous study, 
arXiv:1410.3424 [hep-th],  to rotating strings and BTZ black holes.
As in the non-rotating case, BTZ black holes are generated in the large tension limit, in 
which string back-reaction must be taken into account.
When back-reaction is properly accounted for, collapsing heavy, physical, rotating strings do not generate naked 
singularities but only BTZ black holes, including extremal ones.
The rotating string must contain world-sheet excitations in order to have consistent equations of motion; we describe such additional degrees of freedom implicitly in terms of an effective equation of state. 
\end{abstract}
\end{titlepage}
\newpage

\tableofcontents
\noindent\hrulefill
\bigskip



\section{Introduction}\label{intro}

Three dimensional pure gravity is simpler than its higher-dimensional cousins, since it does not propagate local degrees of freedom.
In Hamiltonian formulation there are six apparent degrees of freedom: three from the spatial part of metric 
and three from conjugate momenta.
These can be eliminated by three constraint equations and gauge transformation generated from them.
Pure gravity in three dimensional anti de Sitter space, AdS$_3$, however, has nontrivial dynamics due to  boundary gravitons~\cite{bh} and black holes~\cite{btz}.

The theory is described by the Einstein-Hilbert action
\beq
S_{EH} = \frac{1}{16 \pi G } \int d^3x \sqrt{-g} \left( R + \frac{2}{l^2} \right).
\eeq{EH}
Although  action (\ref{EH}) admits exact AdS$_3$ as vacuum solution, we are interested in an asymptotically AdS$_3$ space, specified by the following boundary conditions:
\bea
g_{tt} = -\frac{r^2}{l^2} + \mathcal{O}\left( 1 \right), \qquad g_{tr} = \mathcal{O}\left( \frac{1}{r^3} \right), \qquad g_{t \phi} = \mathcal{O}\left( 1 \right), \nonumber\\
g_{rr} = \frac{l^2}{r^2} + \mathcal{O}\left( \frac{1}{r^4} \right), \quad g_{r\phi} = \mathcal{O}\left( \frac{1}{r^3} \right), \quad g_{\phi \phi} = r^2 + \mathcal{O} \left( 1 \right).
\eea{bd}
These boundary conditions are preserved by diffeomorphisms generated by vector fields of the form
\bea
\zeta^t &=& l [ f_+(x^+) + f_-( x^- ) ] + \frac{l^3}{2r^2} [  \partial^2_+ f_+(x^+ ) + \partial^2_- f_-( x^-) ] + \mathcal{O} ( r^{-4}), \nonumber \\
\zeta^{\phi} &=& [f_+(x^+ ) - f_-(x^- ) ]  - \frac{l^2}{2r^2} [ \partial^2_+ f_+(x^+ ) - \partial^2_- f_-(x^-)\ + \mathcal{O} (r^{-4}) , \\
\zeta^r &=& -r [ \partial_+ f_+(x^+) + \partial_- f_-(x^- ) ] + \mathcal{O}(r^{-1}). \nonumber
\eea{diff}
The functions $f_{\pm} ( x^{\pm} )$ are arbitrary and depend only on $x^{\pm} = \frac{t}{l} \pm \phi$.
The time coordinate $t$ and angular coordinate $\phi \sim \phi + 2 \pi$ parametrize the AdS$_3$ boundary at $r = \infty$ and $2\partial_{\pm} = l \partial_t \pm \partial_{\phi}$.

The classical Poisson brackets of the Fourier modes of $f_{\pm}$ form two copies of the Virasoro algebra with equal central charge
$c = 3l/2G$ \cite{bh}.
Therefore, upon quantization, quantum states should fall into unitary representations of the Virasoro algebras.
In modern AdS/CFT context, this suggests that quantum gravity on AdS$_3$ is dual to a two dimensional conformal field theory(CFT) living on the boundary~\cite{mal}.
A boundary CFT appearing naturally in pure AdS$_3$ gravity is  Liouville theory. How such theory arises can be seen both in the second-order Einstein action and in the first-order Chern-Simons formulation. We'll sketch the well
known derivation leading to it in the Chern-Simons language of ref.~\cite{chd}.

The Einstein-Hilbert action (\ref{EH}) is equivalent to two Chern-Simons theories with $SL(2,R)$ gauge group
\bea
S_{EH} &=& S_{CS,k}[A] - S_{CS, k} [\tilde{A}] , \nonumber \\
S_{CS,k} &=& \frac{k}{4 \pi} \mathrm{Tr} \int_{\mathcal{M}} \left( A \wedge dA + \frac{2}{3} A \wedge A \wedge A \right) + \mbox{ boundary terms}.
\eea{CS}
The gauge potentials $A$ and $\tilde{A}$ are related to the dreibein $e^a$ and spin connection $\omega^a$ by
\beq
A = \left( \omega^a + \frac{e^a}{l} \right)T^a ,  \quad \tilde{A} = \left( \omega^a - \frac{e^a}{l} \right) T^a ,
\eeq{connection}
where $T^{1,3}=\sigma^{1,3}$, $T^2=i\sigma^2$. 
By assuming that the three manifold $\mathcal{M}$ has the topology of a solid cylinder and solving the Gauss law
constraints, the gauge potentials 
become $A = g^{-1} d g$ and $\tilde{A} = \tilde{g}^{-1} d \tilde{g} $  for some $g, \tilde{g} \in Sl(2,R)$.
This reduces the Chern-Simons theories in (\ref{CS}) to boundary integrals and they become two copies of chiral WZW theories in the $A_- = \tilde{A}_+ = 0$ gauge.
If the gauge potentials are further restricted to be compatible with  boundary condition~(\ref{bd}), one 
obtains Liouville theory~\cite{chd}.

Black hole solutions compatible with boundary condition (\ref{bd}) also exist~\cite{btz},
\bea
ds^2 &=& -N^2 dt^2 + N^{-2} dr^2 + r^2 ( N^{\phi} dt + d \phi ), \nonumber \\
N^2 &=& -8GM + \frac{r^2}{l^2} + \frac{16G^2 J^2}{r^2} , \quad N^{\phi} = -\frac{4GJ}{r^2}.
\eea{metric}
The entropy of such black holes (BTZ black holes) is given by the Bekenstein-Hawking formula
\beq
S_{BH} = \frac{\mathrm{Area}}{4G} = \frac{2 \pi r_+}{4G}.
\eeq{entropy}
Assuming the existence of a dual CFT, we should be able to reproduce the black hole entropy by counting the number of corresponding states in the dual CFT.
This idea can be checked using Cardy's formula~\cite{cardy}
\beq
d(\Delta , \bar{\Delta}) = \exp \left(2\pi \sqrt{\frac{c \Delta}{6}} + 2\pi \sqrt{\frac{c \bar{\Delta}}{6}} \right),
\eeq{cardyformula}
with the identification~\cite{strom}
\beq
M = \frac{1}{l} (\Delta + \bar{\Delta}) , \quad J = (\Delta - \bar{\Delta}).
\eeq{iden}

The Hamiltonian reduction summarized above justifies studying quantum gravity using Liouville theory, but the
Bekenstein-Hawking and Cardy formulae point  out to a serious obstacle to a naive identification of Liouville
as the CFT dual of pure gravity.  
The Cardy formula (\ref{cardyformula}) assumes a discrete spectrum and the existence of an $SL(2,C)$ invariant vacuum, but these properties do not hold for Liouville theory~\cite{sei}.
By replacing $c$ with $c_{eff} = c - 24 \Delta_0$, where $\Delta_0$ is the smallest conformal weight in the given theory, we can apply the Cardy formula (\ref{cardyformula}) to theories without $SL(2,C)$-invariant vacuum.
But $c_{eff}$ for Liouville theory equals $1$ and this gives too small an entropy compared to the Bekenstein-Hawking formula~ (\ref{entropy}).
Moreover we assumed that spacetime has a trivial topology, but this is no longer true in the presence of black holes. 
Last but not least, we implicitly assumed that the continuous spectrum of the Liouville CFT was discretized 
by some unspecified regularization, since a continuous spectrum yields strictly infinite entropy for states with finite maximum energy.

Nevertheless, there is some positive evidence that pure gravity on AdS$_3$ is strictly related to Liouville theory.
In ref.~\cite{ver}, it was argued that wave functions of Chern-Simons theory with $SL(2,R)$ gauge group are Virasoro conformal blocks.
Due to the equivalence between pure gravity on AdS$_3$ and two copies of $SL(2,R)$ Chern-Simons theory, we expect that the Hilbert space is (schematically) a direct product of two unitary representations of the Virasoro algebra.
We will argue in an upcoming publication that quantum states of pure gravity on AdS$_3$ are one-point functions of conformal field theory with continuous spectrum, bounded below by $\frac{c-1}{24}$~\cite{kimporrati}.

It seems that we have two conflicting arguments about the role of Liouville theory.
The tension between them can be resolved by interpreting Liouville theory as a description of Virasoro descendants~\cite{mart}.
In such case, other information is needed to determine the spectrum of primary fields; in particular, extra degrees of freedom 
must be introduced to complete the theory.
If we are to take seriously Liouville theory as a candidate dual of ``pure'' gravity instead, we must look for special additional degrees of freedom. 
These new \emph{matter degrees of freedom} should be able to approach arbitrarily close to the boundary to account for the continuous
spectrum of Liouville theory and they should be heavy to be decoupled from the low-energy pure gravity sector.
One natural candidate could be long strings~\cite{mawit}. 

String dynamics can be studied in various regimes.
The probe approximation of long strings, with tension $l^{-2} \ll T \ll G^{-1}l^{-1}$, was considered in ref.~\cite{sw}.
This regime is good because it justifies neglecting both the back-reaction of the string and quantum effects. Its drawback 
is that it possesses states with conformal weight well below the Seiberg bound. 
We can also study the $T \ll l^{-2} $ regime via string theory on AdS$_3  \times M$~\cite{gkrs}.
In such a regime strings cannot be heavy so we have to send $g_s \rightarrow 0$ to try to decouple them from pure gravity sector, but this cannot be done in unitary CFTs~\cite{kp}.
In ref.~\cite{kp}, it was argued that $T \gg G^{-1}l^{-1}$ is the relevant regime to study, but in that paper
 only non-rotating strings and BTZ black holes were studied.
The rest of this paper is devoted to generalizing the work of ref.~\cite{kp}  to rotating BTZ black holes.

In section \ref{rotating} we will see that the dynamics of long string can be studied using an effective potential, which
will allow us to  read off some useful information from the string's asymptotic behavior.
In section \ref{large}, we will consider the large tension limit $TGl \gg 1$ and argue that our construction generates only  rotating BTZ black holes when the tension is of the order of the central charge $c$.
In section \ref{small} we will consider the small tension limit and argue that at $TGl \lesssim 1$ our effective 
 description for the long string degrees of freedom breaks down and that a full string description of the 
dynamics may be necessary even at the classical level.

\section{Long Strings and Rotating BTZ}\label{rotating}

Let us consider a collapsing shell of matter approaching from the boundary of AdS$_3$ at past infinity.
The shell with rotational symmetry is a closed string in three dimension, and its worldsheet $\Sigma$ separates the space into two disconnected components $\mathcal{M}_+$ and $\mathcal{M}_-$ with $\Sigma = \partial \mathcal{M}_+ = \partial \mathcal{M}_-$.
From now on, the subscript or superscript $(+)$/$(-)$ will be used for quantities defined in the region outside/inside of the string. 
The system is described by the action
\beq
S = \frac{1}{16 \pi G} \int_{\mathcal{M}_+} d^3x \sqrt{-g_+} \left( R +  \frac{2}{l^2_+} \right) + \frac{1}{16 \pi G} \int_{\mathcal{M}_-} d^3x \sqrt{-g_-} \left( R + \frac{2}{l^2_-} \right) + S_{string}.
\eeq{action}
For now $S_{string}$ could be any kind of action defined on the worldsheet $\Sigma$; in particular, it may include an effective action for other worldsheet  degrees of freedom.
Variation of the action (\ref{action}) with respect to the induced metric $g_{ij}$ on $\Sigma$ gives the Israel junction conditions 
\beq
\gamma^+_{ij} - \gamma^-_{ij} = 8 \pi G S_{ij} , \quad \gamma^{\pm}_{ij} \equiv K^{\pm}_{ij} - g_{ij} K^{\pm} ,
\eeq{junction}
where $S_{ij}$ is the energy-momentum tensor of the string and $K^{\pm}_{ij}$ denote extrinsic curvatures.

In the regime $T \gtrsim G^{-1}l^{-1}$ the back-reaction of the string must be considered, so the  metric can change across the string worldsheet $\Sigma$.
Let us assume that we have a rotating black hole metric outside and pure AdS$_3$ metric inside:
\bea
ds^2_- &=& -N^2_- dt^2 + N^{-2}_- dr^r + r^2 d{\phi}^2 , \nonumber \\
ds^2_+ &=& -N^2_+ dt^2 + N^{-2}_+ dr^2 + r^2 \left( N^{\phi} dt + d\phi \right)^2 ,  \\
N^2_- &=& 1 + \frac{r^2}{l^2_-} ,  \quad N^2_+ = -8GM + \frac{r^2}{l^2_+} + \frac{16 G^2 J^2 }{r^2} , \quad  N^{\phi} = - \frac{4GJ}{r^2}. \nonumber
\eea{gluing}
To apply the Israel junction conditions we must know the metric $g_{ij}$ on the string worldsheet $\Sigma$, but a naive projection of the metrics $g_+$ and $g_-$ onto $\Sigma$ would not work, due to the discontinuity of some $g_{t\phi}$ components.
Instead, we can choose a rotating frame for the region outside the string, spanned by
\beq
e^{\mu}_{\tau} = \left( \dot{T}_+ , \quad \dot{R} , \quad \dot{\Phi} \right) , \qquad e^{\mu}_{\theta} = \left( 0 , \quad 0, \quad \frac{1}{R} \right).
\eeq{frame}
The radial coordinate $R$ is the radius of the string and $\dot{X}$ denotes the derivative of $X$ with respect to $\tau$.
Although the coordinate radial velocity $V_R = \frac{\dot{R}}{\dot{T}_+}$ should be the same as that of the string, the coordinate angular velocity $\Omega_+ = \frac{\dot{\Phi}}{\dot{T}_+}$ differs from that of the string in general.
By setting $\Omega_+ = \frac{4GJ}{R^2}$, one can go to a convenient local frame, i.e. $g_{ij} = \mathrm{diag} ( -1 , 1)$, and the normalization condition of $e_{\tau}$ reads $N^2_+ \dot{T}^2_+ - N^{-2}_+ \dot{R}^2 = 1$.

Using these conditions it is straightforward to compute $\gamma^+_{ij}$:
\bea
\gamma^+_{\tau \tau} ~=~ -\frac{\beta_+}{R} , \quad  \gamma^+_{\tau \theta} &=& - \frac{4GJ}{R^2} , \quad \gamma^+_{\theta \theta} ~=~ \frac{d}{dR} \beta_+ , \nonumber\\
\beta_+ &=& \sqrt{ N^2_+ ~+~ \dot{R}^2}.
\eea{gammaplus}
We can immediately obtain $\gamma^-_{ij}$ by setting $J=0$ and replacing $\beta_+$ with $\beta_- =\sqrt{ N^2_- + \dot{R}^2}$; therefore,
$\gamma_{ij} \equiv \gamma^+_{ij} - \gamma^-_{ij} $ is given by
\beq
\gamma_{\tau \tau} = -\frac{1}{R} \left( \beta_+ - \beta_- \right) , \quad \gamma_{\tau \theta} = -\frac{4GJ}{R^2} , \quad \gamma_{\theta \theta} = \frac{d}{dR} \left( \beta_+ - \beta_- \right).
\eeq{gamma}
Note that the non-rotating case can be obtained by setting $J=0$ in the above equations.

A nonzero $\gamma_{\tau \theta}$ is inconsistent with the equation of state $\rho = -p = T$ for a string tension $T$ since such equation of state gives $S^i_j = - T \delta^i_j$ in any coordinate system.
One possible generalization is to excite some additional massless degrees of freedom on the string worldsheet, so that the equations of state read
\beq
\rho = T + \epsilon , \qquad p = - T + \epsilon .
\eeq{ansatz}
The excitation energy $\epsilon$ does not change the trace of energy-momentum tensor 
so we avoid possible issues with the  trace anomaly.
The need for the excitation energy $\epsilon$ can be understood because a nonzero spacetime angular momentum of the string in AdS$_3$ can be achieved by giving different excitation to left and right movers \cite{malo}.
This means that we need some as yet unspecified additional degrees of freedom on the string worldsheet. We do
not need to know the microscopic description of such degrees of freedom in this work, so we will use an 
effective description given by equations of state~(\ref{ansatz}). 

The energy-momentum tensor $S_{ij}$ can be computed by specifying $U_i$, the projection of the three-velocity $U^{\mu}=\left( {T}^{\prime}_+ , {R}^{\prime}, \Phi^{\prime} \right)$ onto $\Sigma$.
The three-velocity should be measured by following a fixed point on the string. Its corresponding proper time, say $\tau_s$, could be different from $\tau$, the proper time of the rotating frame (\ref{frame}).
The prime denotes the derivative with respect to $\tau_s$.
As mentioned earlier we have $V_R = \frac{\dot{R}}{\dot{T}_+} = \frac{R^{\prime}}{T^{\prime}_+}$ but $\Omega_+ \neq \Omega_s \equiv \frac{\Phi^{\prime}}{T^{\prime}_+}$.
It is useful to define
\beq
V_{\Phi} = R \dot{T}_+ \left( \Omega_s - \Omega_+ \right),
\eeq{angular}
which can be interpreted as the angular component of the string velocity, measured in the local frame spanned by vectors (\ref{frame}).

We finally obtain the energy-momentum tensor for the string using the normalization condition for $U^{\mu}$:
\beq
S_{\tau \tau} = T + \left( \frac{1 + V^2_{\Phi}}{1- V^2_{\Phi}} \right) \epsilon , \quad S_{\tau \theta} = - \left( \frac{2 V_{\Phi}}{1 - V^2_{\Phi}} \right) \epsilon , \quad S_{\theta \theta} = - T + \left( \frac{ 1 + V^2_{\Phi}}{1 - V^2_{\Phi}} \right)\epsilon .
\eeq{stess}

The  junction conditions now read
\bea	
-\frac{1}{R} \left( \beta_+ - \beta_- \right) & = &  8\pi G T + F(R) , \nonumber \\
\frac{d}{dR} \left(\beta_+ - \beta_- \right) & = & - 8 \pi G T  + F(R) , \nonumber \\
F(R) & = & \left( \frac{1 + V^2_{\Phi}}{ 2 V_{\Phi}} \right) \Omega_+ . 
\eea{eom}
The existence of a solution forces $F(R)$ to be proportional to $R^{-2}$; this implies that $V_{\Phi}$ is a constant.
Another check of this result comes from conservation of $S_{ij}$:
\beq
\nabla_i {S^i}_j = \partial_i {S^i}_j   ~+~ {{\omega_i}^i}_k {S^k}_j ~-~ {{\omega_i}^k}_j {S^i}_k = 0.
\eeq{conservation}
Using the zero-torsion conditions $de^i + {\omega^i}_j \wedge e^j = 0$, it is easy to see that the only non-vanishing component of the spin connection is ${{\omega_{\theta}}^{\theta}}_{\tau} = \frac{\dot{R}}{R}$.
By plugging this result into the conservation equation (\ref{conservation}) one can show that
\beq
\frac{ \left( 1 \pm V_{\Phi} \right)^2}{1 - V^2_{\Phi}}  \epsilon = \frac{C_{\pm}}{R^2},
\eeq{consistency}
for some constants $C_{\pm}$.
This implies that $V_{\Phi}$ is a constant, confirming the above result. It also allows us to express $\epsilon$ as
\beq
\epsilon  = \left( \frac{ 1 - V^2_{\Phi}}{ 2 V_{\Phi}} \right) \frac{J}{2\pi R^2} .
\eeq{epsilon}

The Israel junction condition (\ref{eom}) can be easily solved to give an equivalent one dimensional problem of a point particle with zero energy~\cite{mo}.
\bea
\dot{R}^2 + V_{eff} (R) &=& 0  , \nonumber \\
V_{eff} (R) &=& \frac{4 N^2_+ N^2_- - \left[ N^2_+ + N^2_- - R^2 \left( 8\pi GT + \alpha \frac{4GJ}{R^2} \right)^2 \right]^2}{4R^2 \left( 8\pi GT +  \alpha \frac{4GJ}{R^2} \right)^2} , \\
\alpha &=& \frac{1 + V^2_{\Phi}}{2 V_{\Phi}} . \nonumber
\eea{effective}
The unitless constant $\alpha$ diverges as $V_{\Phi} \rightarrow 0$ and it goes to $\pm 1$ as $V_{\Phi} \rightarrow \pm 1$.
{So we will keep $\alpha \neq \pm 1$ to ensure that  the motion of the string is subluminal.}
The positivity of $\epsilon$ means that all of $J$, $\alpha$, and $V_{\Phi}$ have the same sign, so we may restrict ourselves
to the $J>0$, $\alpha > 1$ (or equivalently $V_{\Phi} > 0$ ) case whenever it is necessary to specify the sign of $J$.

It is also interesting to examine the asymptotic behavior of the string.
Since the string must be able to approach the AdS boundary with finite speed, the asymptotic form of $\dot{R}^2$ should be
\beq
\dot{R}^2 = a^{(0)} + \frac{a^{(-2)}}{R^2} + \mathcal{O} \left( \frac{1}{R^4} \right).
\eeq{Rsquare}
Let us plug this condition  into the junction condition (\ref{eom}) and collect terms of the same order in $R$.
Then, the leading order terms express the string tension in terms of $l_+$ and $l_-$:
\beq
8\pi G T = \frac{1}{l_-} - \frac{1}{l_+} .
\eeq{leading}
This gives the same critical tension introduced in ref.~\cite{kp}.
If this condition is not met,  the string either cannot approach the boundary or it reaches it with infinite speed (in the coordinate $R$).
Therefore, from now on we will focus on strings with critical tension.
The subleading order terms in~(\ref{eom}) give a relation between $(M, J)$ and $(a^{(0)} , \alpha , l_+ , l_-)$. 
It is useful to eliminate $l_-$ in favor of the reduced tension $\tilde{T} \equiv 8\pi GT l_+$ and to measure the 
black hole  mass $M$ in AdS unit $l_+$:
\beq
M l_+ = - \left( \frac{c}{12} \right) \left( \frac{1}{ 1+ \tilde{T}} \right) + a^{(0)} \left( \frac{c}{12} \right) \left( \frac{\tilde{T}}{1+\tilde{T}} \right) + \alpha J .
\eeq{subleading}

The black hole mass $M$ and angular 
momentum $J$ can be determined once the dynamical and kinematical variables of the string are known.
From the definition of $V_{\Phi}$ and the fact that it is a constant, one also obtains $V_{\Phi} = \Omega^{(0)}_s l_+$, where $\Omega^{(0)}_s$ is defined as the asymptotic value of $\Omega_s$ in the same way as we defined $a^{(0)}$.
So the angular momentum $J$ can be determined by specifying the energy excitation $\epsilon$ above the vacuum,
while the mass $M$ is obtained by knowing $a^{(0)}$:
\beq
M = M\left(T, \epsilon , a^{(0)} , \Omega^{(0)}_s ; l_+ \right) , \qquad J = J \left( \epsilon ,\Omega^{(0)}_s ; l_+ \right).
\eeq{MJ}
Notice that the angular momentum $J$ does not depend on $a^{(0)}$.

\section{Large Tension Limit}\label{large}

In ref.~\cite{kp} we argued that long strings give a continuous spectrum with a correct lower bound for black hole masses in the large tension limit.
More precisely, we found that the Seiberg bound is obeyed in the $ TGl_+ \sim c$ regime. 
The same regime also guarantees the 
decoupling of long strings from the pure gravity sector and implies that the space bounded  by the string worldsheet 
has Planckian curvature, i.e. $ l_- \sim G$.
It is therefore reasonable to look at the same limit for rotating cases.

Let us start by looking at the infinite tension limit. There the subleading term in~(\ref{subleading}), proportional to $c/\tilde{T}$, approaches zero from below and the leading term could be any positive number since $a^{(0)} \geq 0$ is the only restriction.
Therefore we have
\beq
M l_+ \gtrsim \alpha J .
\eeq{massbound}
The positivity of $\alpha J$ assures that black hole mass is non-negative and $\vert J \vert \leq M l_+$ since $\vert \alpha \vert > 1$.

This looks promising but one must ask what happens when the tension is large but finite.
In this case we have to make sure that the $c/\tilde{T}$ term does not produce any negative mass black holes, since we cannot simply neglect it.
It is particularly interesting to look at the $\tilde{T} \sim c$ regime, which implies $l_- \sim G$.
In this regime $c/\tilde{T}$ becomes order of unity but it can be easily canceled by a $ c a^{(0)}$ term.
So we recover $M l_+ \gtrsim \alpha J$ and we have a correct black hole spectrum which includes extremal black holes ($M l _+ =J$).
Note that requiring the string to move with subluminal angular velocity implies that 
extremal black holes occur only when $a^{(0)} < \frac{1}{\tilde{T}}$.

In the presence of a nonzero angular momentum $J$, one may expect a turning point in the motion of the collapsing 
long strings, due to the centrifugal barrier, but for finite back-reaction  such turning point may be hidden by an event horizon.
More precisely we want to ask the following question: 
once we specify the mass and the angular momentum of a black hole, can we find a long string whose turning points are hidden by the outer horizon?
It turns out that we can find at least one such long string for any black hole.
This can be seen by studying zeros of the numerator of the effective potential, which we shall call  $V^{num}_{eff}$.

For this purpose it is convenient to work with a cubic equation $V(y) \equiv y^2 V^{num}_{eff} ( y) =0 $, where $y \equiv R^2$, since it has the same solutions as $V^{num}_{eff} = 0 $ except the $y = 0 , \infty$ ones.
By eliminating $l_-$ in favor of the reduced tension $\tilde{T}$ we have
\beq
V(y) =  A y^3 + By^2 + Cy + D  ,
\eeq{numerator}
where
\bea
A &=& -  \left[ \frac{4 a^{(0)}}{l^2_+}  \right] \tilde{T}^2  ,  \nonumber \\
B &=& \left[ -\left({\alpha}^2 -1 \right) \left(\frac{8GJ}{l_+}\right)^2 \right]  \tilde{T}^2 +  \left[ 2 \alpha \left(1 - 8GM \right)  + \left( {\alpha}^2 + 1 \right) \left(\frac{8GJ}{l_+}\right)  \right] \left( \frac{8GJ}{l_+} \right) \tilde{T},  \nonumber \\
  & &+ \left[ - \left( 1+ 8GM \right)^2  + {\alpha}^2 \left(\frac{8GJ}{l_+} \right)^2 \right] , \nonumber \\
C &=& \left[ - \alpha \left( {\alpha}^2 - 1 \right)  \left(\frac{8GJ}{l_+} \right) \tilde{T}  +  \left[ \left( {\alpha}^2 +1 \right) - 8GM \left(  {\alpha}^2 - 1 \right) \right] \right] \left( \frac{l^2_+}{2} \right) \left(\frac{8GJ}{l_+}\right)^2 ,\nonumber \\
D &=& -\left({\alpha}^2 - 1 \right)^2 \left( \frac{l^4_+}{16} \right) \left( \frac{8GJ}{l_+} \right)^4 . 
\eea{coeff}

Let us start with an extremal black hole ($M l_+ = J$) with $ \alpha = a^{(0)} \tilde{T} = 1$.
The physical relevance of this case is unclear yet, because the string rotates with the speed of light when $\alpha = 1$, 
but this is not a problem, since we will show that the physical $\alpha>1$ case can be studied by performing a small perturbation around $\alpha=1$. 
When $\alpha = a^{(0)} \tilde{T} = 1$, the cubic equation $V(y) = 0$ takes the very simple form
\beq
-\frac{4 \tilde{T}}{l^2_+} y^3 + \left( 16GM \tilde{T} - \left( 1 + 16GM \right) \right) y^2  + 64G^2 M^2 l^2_+ y = 0,
\eeq{extreme}
and its nonzero solutions are
\bea
y_{\pm} &=& \left( \frac{16 GM \tilde{T} - \left( 1 + 16GM \right) \pm \sqrt{\Delta} }{8\tilde{T}} \right) l^2_+ , \nonumber \\
\Delta &=& \left(16 GM \tilde{T} - \left(1 + 16GM \right) \right)^2 +  1024 M^2 G^2 \tilde{T} .
\eea{sol}
One can easily see that $y_-$ is irrelevant because it is always negative,
and $y_+$ is positive but smaller than the square of the outer horizon, $r^2_+ = 4GM l^2_+$, for any value of the black hole mass $M$ and the tension $\tilde{T}$.

\begin{figure}
	\begin{center}
	  \subfloat[$\tilde{T}  = 100,~~\alpha = 1.0001$]{
	    \includegraphics[width=0.47\linewidth]{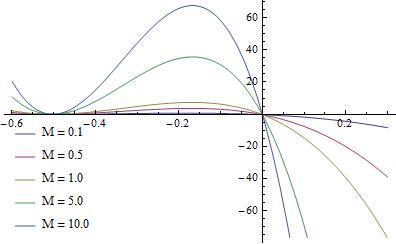}}
	  \subfloat[$\tilde{T}  = 100,~~\alpha = 1.0001$]{
	    \includegraphics[width=0.47\linewidth]{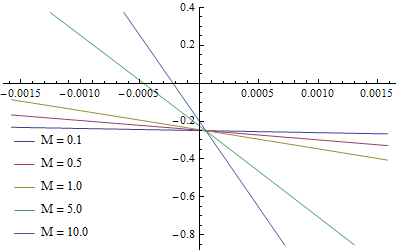}}
	    \\
	  \subfloat[$\tilde{T}  = 1000,~~\alpha = 1.00001$]{
	    \includegraphics[width=0.47\linewidth]{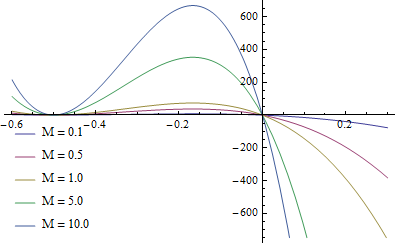}}
	  \subfloat[$\tilde{T}  = 1000,~~\alpha = 1.00001$]{
	    \includegraphics[width=0.47\linewidth]{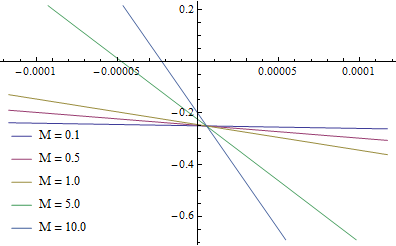}}
	    \\
	  \subfloat[$\tilde{T}  = 7500,~~\alpha = 1.00001$]{
	    \includegraphics[width=0.47\linewidth]{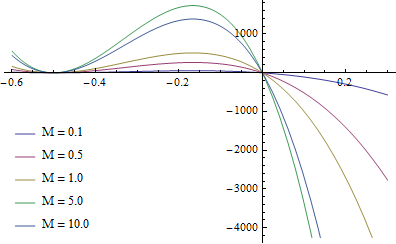}}
	  \subfloat[$\tilde{T}  = 7500,~~\alpha = 1.00001$]{
	    \includegraphics[width=0.47\linewidth]{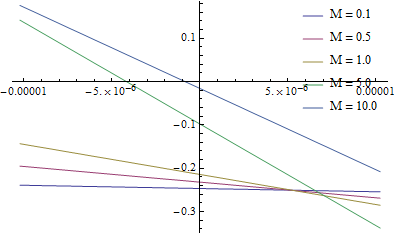}}
		
	  \caption{We redefine the argument of $V(y)$ as $\left( y+\frac{1}{2} \right)(M l^2_+)$ so that the horizon is sitting at the origin, and we set $8G = 1$.
		From figure (a), (c) and (e), one can see that the largest turning points occur near the horizon (the origin here).
		Figure (b), (d) and (f) zoom in near the origin to show that the turning points are hidden behind the horizon.}
		\label{plot_hidden}
	\end{center}
\end{figure}

Now let us consider how $y_+$ changes by a small perturbation in $\alpha$, or equivalently in $a^{(0)} \tilde{T}$.
From the turning point condition $V(y_+ ) = 0$, we have
\beq
\left( \left. {\frac{\partial V}{\partial \alpha} }\right\vert_{\begin{smallmatrix} \alpha = 1 \\ y_+ \end{smallmatrix}}  \right)
+ \left( \left. {\frac{\partial V}{\partial y}} \right\vert_{\begin{smallmatrix} \alpha = 1 \\ y_+ \end{smallmatrix}}  \right)
\left( \left. {\frac{dy_+}{d \alpha}} \right\vert_{ \alpha = 1 }  \right) = 0.
\eeq{pertur}
Using the equality of the subleading order terms in the asymptotic behavior (\ref{subleading}), the cubic equation (\ref{extreme}) and its solution (\ref{sol}), one can obtain
\beq
y^{\prime} \equiv \left. {\frac{dy_+}{d  \left( a^{(0)} \tilde{T} \right)  }} \right\vert_{\begin{smallmatrix} \alpha =1 \\ y_+ \end{smallmatrix}}
= - \frac{1}{\tilde{T} + 1} \left( \frac{ \left(\tilde{T} - 1 \right) y_+ + 8GMl^2_+ }{\sqrt{\Delta}} \right)  < 0.
\eeq{dyda}
This means that the turning point increases as we decrease $a^{(0)} \tilde{T}$.
Since we want to hide all turning points behind the outer horizon, the maximum allowed change of $a^{(0)} \tilde{T}$, say $\delta_{max}$, is approximately 
\beq
\delta_{max} \approx  \frac{ r^2_+ - y_+}{\left\vert y^{\prime} \right\vert}  .
\eeq{delta_max}
In general $\delta_{max}$ could be very small but it is nonvanishing for finite black hole mass $M$ and tension $\tilde{T}$.
Therefore for any extremal black hole, we can find a long string with $a^{(0)} \tilde{T} \in ( 1 - \delta_{max}, 1)$ such that all turning points are hidden by the outer horizon.\footnote{
The approximate expression (\ref{delta_max}) does not make sense if the second derivative $y^{\prime \prime} \equiv \frac{d^2 y_+}{d \left( a^{(0)}  \tilde{T} \right)^2} $ diverges since the error of the approximation (\ref{delta_max}) depends on it.
However the finiteness of $y^{\prime \prime}$ follows the facts that $V$ is analytic in $\alpha$ and $\frac{\partial V}{\partial y}$ is nonvanishing at $y_+$.}
The behavior of $V(y)$ for various values of $\tilde{T}$ and $M$ with $\alpha \neq 1$ is plotted in figure (\ref{plot_hidden}) to show that the turning points are hidden behind the horizon.

The generalization to the non-extremal case can be done by decreasing the angular momentum $J$ from the extremal 
value $|J|=Ml_+$.
As we decrease the angular momentum $J$, the centrifugal barrier will weaken, resulting in a smaller value for $y_+$.
This can be checked by changing $\alpha$ and $J$ while keeping $ \alpha J$ fixed.
Then the variation of $V$ with respect to $J$ is
\beq
\frac{ \delta V}{\delta J} = \frac{ 128 G^2 J}{l^2_+} \tilde{T} \left( \tilde{T} + 1 \right) y^2 +   64 G^2 J \left(\alpha \frac{ 8GJ}{l_+} \tilde{T} + \left( 1 + 8GM \right) \right) y + 2G \left( \alpha^2 - 1 \right)  \left( 8GJ \right)^3 ,
\eeq{varVJ}
which is clearly positive.
Since the coefficient of the cubic term in $V(y)$ is negative, $\frac{\partial V}{\partial y} $ is negative at the largest turning point $y_+$.
So the variation of $y_+$ with respect to the angular momentum $J$ is
\beq
\frac{\delta y_+}{\delta J} = - \left( \left. \left. \frac{\delta V }{\delta J} \right\vert_{y_+} \right) \right/  \left( \left. \frac{\partial V}{\partial y} \right\vert_{y_+} \right) > 0.
\eeq{varyJ}
On the other hand the radius of the outer horizon of a non-extremal black hole is larger than that of extremal one.
This completes the proof of the existence of a long string state with turning points hidden behind the horizon
for a black hole with arbitrary mass $M$ and angular momentum $J$.

One interesting observation regarding turning points is that rotating strings with fixed radius can create extremal black holes.
This can be seen by expressing $V^{num}_{eff}(y)$ in terms of $z = \frac{y}{8GMl^2_+} - \frac{1}{2}$ so that the horizons of extremal black holes are at the origin.
Then turning points can be studied with the help of the cubic function
\beq
\tilde{V}(z) \equiv z^2 V^{num}_{eff} \left( y(z) \right) = \tilde{A} z^3 + \tilde{B} z^2 + \tilde{C} z + \tilde{D}, 
\eeq{newpotential}
where,
\bea
\tilde{A} &=& 32GM \left[ 8(\alpha - 1)GM\tilde{T} + 8(\alpha -1)GM -1 \right] \tilde{T}  , \nonumber \\
\tilde{B} &=& -64(\alpha - 1 )(\alpha -5) G^2 M^2 \tilde{T}^2 
			  + \left[ 64 (\alpha - 1)(\alpha + 5) G^2M^2 + 16 (\alpha -3 )GM 		\right] \tilde{T} \nonumber \\
		  & & - ( 1 + 8GM)^2 + 64 \alpha^2 G^2 M^2 , \nonumber \\
\tilde{C} &=& \left[ -8(\alpha -2) GM \tilde{T} + 8GM  + 1 \right]
			  \left[ 8(\alpha - 1) GM \tilde{T} + 4 (\alpha^2 - 1) GM - 1 \right] , \nonumber \\
\tilde{D} &=& -\frac{1}{4} \left[ 8(\alpha - 1)GM \tilde{T} + 4(\alpha^2 - 1)GM  - 1 \right]^2 .
\eea{newcoefficeint}  
Here we eliminated $a^{(0)}$ in terms of $M$, $J$, $\tilde{T}$ and $\alpha$ using relation (\ref{subleading}).

Obviously the $\tilde{D}$ term is negative unless
\beq
\alpha = \alpha^* \equiv - \tilde{T} + \sqrt{ \left( \tilde{T} + 1 \right)^2 + \frac{1}{4GM} } .
\eeq{static}
When this condition is met, both $\tilde{C}$ and $\tilde{D}$ vanish but $\tilde{B}$ does not.\footnote{
The parameter $\alpha^*$ is obviously greater than 1, and the other value for $\alpha$ to give $\tilde{C} = 0$ is 
greater than $\alpha^*$.
Also $\tilde{B}$ is negative for $\alpha \in (1 , \alpha^*)$.}
This implies that $\tilde{V}(0) = \tilde{V}^{\prime}(0) = 0$ and $\tilde{V}^{\prime \prime}(0) \neq 0$ when $\alpha = \alpha^*$, so $\tilde{V}(z)$ has one negative root and one degenerate root at zero.
Therefore if a rotating string is at $r = \sqrt{ 4GM} l_+$, an extremal black hole with mass $M$ can be created.
When $\alpha$ is greater than $\alpha^*$ on the other hand, $\tilde{V}^{\prime}(0)$ is positive since $\tilde{C}$ is positive.
The turning points are not hidden behind the horizon in these cases, so $\alpha^*$ is the upper bound in $\alpha$ to have hidden turning points.

We want to conclude this section by considering the scale of energy excitations above the vacuum.
The energy density $S_{\tau \tau}$ consists of the vacuum energy density $T$ and the excitation energy $\Delta_{\rho} = \frac{\alpha J}{2\pi R^2}$.
Since $S_{ij}$ is only the energy-momentum tensor of an effective theory, we can only expect it to be reliable when $\Delta_{\rho} /T \ll 1$ holds outside the outer horizon $r_+ \sim \sqrt{GM} l_+$. This is guaranteed in the 
$\tilde{T} \sim c$ regime.

Thus, it seems that $\tilde{T} \sim c$ is the correct regime to consider: 
it gives the correct spectrum for black hole masses and angular momenta;
we can find a long string state which generates all the future outside region of any given  BTZ black hole, because the
would-be  turning point of the collapsing string are hidden by the BTZ outer horizon; finally, 
the excitation energy is very small compared to the vacuum energy.
However, we cannot rule out other regimes.
For example, we can consider the $1 \ll \tilde{T} \ll c$ regime.
A correct spectrum can still be obtained by properly restricting $a^{(0)}$.
It may require more computations but the energy excitation can still be small and a string with hidden turning 
points can still be found for any black hole.

\section{Small Tension Limit}\label{small}

String theory on AdS$_3$ has been studied extensively, see for example \cite{malo, gks}, so it is worth to find connections with the long string studied in this paper.
In exact string theory calculations the background metric is fixed, 
so we are within an approximation where the tension should be small enough to neglect back-reaction, $TGl \ll 1$.
On the other hand, our analysis was confined to the classical level, so the tension should be large enough to neglect quantum effect, $l^{-2} \ll T$.
Therefore we are in the probe approximation considered in \cite{sw}.

The first thing we observe is the appearance of negative mass 
states~\footnote{We recall that here $M=0$ is the minimum BTZ mass.}.
In this regime the subleading order equality (\ref{subleading}) reads
\beq
Ml_+ = -\frac{c}{12} + \frac{c}{12} \left( 1 + a^{(0)} \right) \tilde{T} + \alpha J + \mathcal{O} \left( {\tilde{T}}^2 \right) . 
\eeq{smalltension}
The above equation and the fact that $c$ is very large imply that we have a large number of negative mass states and/or state with $\vert J \vert \geq \vert M \vert l_+$.
One may argue that those states can be removed by adjusting $a^{(0)}$ and $\alpha J$ as we did for the large but finite tension regime.
The difference is that we need more severe restriction in this case: for example, a wider range of $a^{(0)}$ should be forbidden for fixed $\alpha J$.
However, the existence of negative mass states is not a problem in itself.
Indeed we know that there are states below the Seiberg bound in the probe approximation \cite{sw, kp}.
It seems more natural to allow them rather than restricting $a^{(0)}$ without further justification.

We can also study turning points of a long string in the small tension limit.
One may say that, as far as turning points are concerned, there is no clear distinction between the large tension limit and the small tension limit, because the analysis of the previous section is valid for arbitrary tension,
and we can find a long string with hidden turning points for any black hole.
This is not a surprise because the important point here is that we have a black hole geometry even though the tension is small; therefore, gravitational back-reaction cannot be neglected.
The difference from the large tension limit can be found by studying negative mass states.
Here we are especially interested in the motions of long strings that do not appreciably back-react on the geometry.
We expect that such motions have turning points as long as the angular momentum is not zero~\cite{malo}.\footnote{ 
One can actually show that any negative mass state with nonzero angular momentum has a turning point.
However, we do not discuss the general case here, because the proof is more complicated than for very small back-reaction case, and because there is no fixed-background calculation to compare our result with.}
It is important to make clear which limit we are working with.
For the outside metric to be close to the pure AdS$_3$ metric, the mass $M$ goes to $-1$ and the angular momentum goes to $0$.
Since the mass $M$ and the angular momentum $J$ always appear multiplied by the Newton's constant $G$, 
we may get a correct limit by sending $G$ to zero.
Indeed this is the correct limiting procedure for decoupling strings from the gravity sector.
It is also reasonable to think that the difference between $l_+$ and $l_-$ is order of $G$, which implies $\tilde{T}$ scales as $G$.
Let us redefine parameters to make this limit apparent:
\beq
8GM = -1 + mG , \qquad \frac{8GJ}{l_+} = j G , \qquad \tilde{T} = \chi G .
\eeq{Gcoeff}
Here the values of $m$, $j$ and $\chi$ are held fixed and ${\cal O}(1)$.
Then the relation between parameters (\ref{subleading}) tells us that $\alpha$ is ${\cal O}(1)$ and $a^{(0)}$ is of order  $G^n$ with a non-negative $n$.

Turning points can be studied by a cubic function as usual:
\beq
\hat{V}(w) \equiv w^2 V^{num}_{eff} \left( y(w) \right) = \hat{A} w^3 + \hat{B} w^2 + \hat{C} w + \hat{D}, 
\eeq{pot_neg}
where $ y = w l^2_+$ and the coefficients are
\bea
\hat{A} &=& -4 a^{(0)} \chi^2 G^2 ,  \nonumber \\
\hat{B} &=& \left( -m^2 + \alpha^2 j^2 + 4 \alpha j \chi \right) G^2  + \mathcal{O} \left(G^3 \right)  , \nonumber \\
\hat{C} &=& \alpha^2 j^2 G^2 + \mathcal{O} \left(G^3 \right) ,  \nonumber \\
\hat{D} &=& -\left( \alpha^2 - 1 \right)^2 \frac{j^4}{16} G^4 .
\eea{coeff_neg}
If $\hat{V}(w)$ is positive at least for one positive value of $w$, 
we will a have turning point because $V(w)$ goes to $-\infty$ as $w$ goes to $0$ or $+\infty$.
So we only have to check the sign of $\hat{V}(w)$ at the maximum.
The maximum occurs at $w^* = -\left( \frac{ \hat{B} + \sqrt{ \hat{B}^2 - 3 \hat{A} \hat{C} } }{3 \hat{A} } \right) > 0$, and it is
\bea
\hat{V}(w^*) &=& \frac{ \mathcal{F} }{ 27 \hat{A}^2}  , \nonumber \\
\mathcal{F} &=& 2 \left( \hat{B}^2 - 3 \hat{A} \hat{C} \right)^{3/2} + 2 \hat{B}^3 - 9 \hat{A} \hat{B} \hat{C} + 27 \hat{A}^2 \hat{D} .
\eea{maxi_neg}
{For $\hat{B}<0$, $\mathcal{F}$ has the form $\mathcal{F}= X-Y$ with $X,Y$ positive so its sign is the same as $X^2-Y^2$ i.e. }
\bea
\mathrm{Sign} \left[ \mathcal{F} \right] &=& \mathrm{Sign} \left[ 4 \left(\hat{B}^2 - 3 \hat{A} \hat{C} \right)^3 - \left( 2 \hat{B}^3 - 9 \hat{A} \hat{B} \hat{C} + 27 \hat{A}^2 \hat{D} \right)^2 \right] \nonumber \\
 &=& \mathrm{Sign} \left[ \hat{B}^2 \hat{C}^2 - 4\hat{A} \hat{C}^3  - 27 \hat{A}^2 \hat{D}^2  - 4 \hat{B}^3 \hat{D} + 18 \hat{A} \hat{B} \hat{C} \hat{D} \right] .
\eea{sign}
Here $\left( \hat{B}^2 \hat{C}^2 \right) $ and $\left( - 4\hat{A} \hat{C}^3 \right)$ are always 
positive and the others may negative.
The ratios between positive terms and negative terms are
\bea
\frac{ \vert -27 \hat{A}^2 \hat{D}^2 \vert }{ \vert - 4 \hat{A} \hat{C}^3  \vert }  & \sim & \vert \hat{A} \vert G^2 \sim G^{m_1}  , \nonumber \\
\frac{  \vert 18 \hat{A} \hat{B} \hat{C} \hat{D} \vert}{ \vert -4 \hat{A} \hat{C}^3 \vert } & \sim & \vert \hat{B} \vert \sim G^{m_2}  , \nonumber \\
\frac{ \vert -4 \hat{B}^3 \hat{D} \vert }{\vert \hat{B}^2 \hat{C}^2 \vert } & \sim & \vert \hat{B} \vert \sim G^{m_2},
\eea{ratios}
where $m_1$ is greater than or equal to $4$, and $m_2$ is greater than or equal to $2$.
Therefore, the positive terms become dominant in the limit $G\rightarrow 0$, thus the sign of $\mathcal{F}$, equivalently the sign of $\hat{V}(w^*)$, is positive.
For positive $\hat{B}$ the only negative term in ${\cal F}$ is $\left( 27 \hat{A}^2 \hat{D} \right)$, which is always of higher order in $G$ than the $\left( \hat{B}^2 - 3 \hat{A} \hat{C} \right)^{3/2}$ term
\beq
\frac{ \vert 27 \hat{A}^2 \hat{D} \vert }{ \left( \hat{B}^2 - 3\hat{A} \hat{C} \right)^{3/2} }< \frac{ \vert  27 \hat{A}^2 \hat{D} \vert }{ \left( -3\hat{A}\hat{C} \right)^{3/2}} \sim G^{m_2}.
\eeq{ratio2}
Here $m_2\geq 2$, thus $\hat{V} (w^*)$ is positive and the motion of strings has always turning points as long as $j$ is not zero.

Despite agreement with known results, our construction may not give a correct description at quantitative level 
{when back-reaction is non negligible.}
Indeed in the $l^{-2} \ll T \ll G^{-1}l^{-1}$ regime, by setting $\alpha J \sim Ml_+$ for positive mass states, the energy excitation $\Delta_{\rho}$ is much lager than the vacuum energy density $T$ in the near horizon region, so the ansatz (\ref{ansatz}) may break down there.
If we require $\Delta_{\rho} \lesssim T$ outside the event horizon, the string tension should be $ \tilde{T} \gtrsim 1$.
This suggests that there could be a change in gravity theory near $\tilde{T} \sim 1$: if $1 \ll \tilde{T}$ the classical dynamics of long strings gives a good description of the geometry of spacetime, but if $l^{-2} \ll T \ll G^{-1} l^{-1}$ we have to work with full string theory even for a classical analysis. 
With the methods used in this paper, we cannot tell whether the change occurs smoothly or through a phase transition.

\section{Conclusions}\label{conclusion}

In this paper we studied long string dynamics in AdS$_3$ as a first attempt to gain a better understanding of rotating BTZ black holes.
We showed that the large tension limit gives a correct black hole spectrum and that long strings can be decoupled from the pure gravity sector.
We focused on the $TGl \sim c$ regime and argued that our construction should be considered only as 
an effective description.
We showed that we can find a long string with hidden turning points for any given black hole.

A more direct connection with string theory on AdS$_3$ can be made in the small tension limit $l^{-2} \ll \ T \ll G^{-1} l^{-1}$, which gives a correct qualitative description.
But a quantitative analysis should not be taken at face value, 
 since the energy of excitations above the vacuum could be greater than the vacuum energy itself, 
 thereby rendering the validity of our effective description questionable {when the string has enough energy to 
 create a black hole}.
So, near $TGl \sim 1$, our effective equation of state description of dynamics may break down and a complete string 
theoretical description may be necessary even at the classical level for $TGl \lesssim 1$.

To broaden the applicability of our previous analysis~\cite{kp}, we added by hand some excitation energy $\epsilon$
  and this introduced an additional parameter $\alpha$, which gives some difficulty for an analytic study in a general setting.
On the other hand this gives some freedom to our analysis.
For example, one can study an exotic case by setting $\alpha = 1 $.
In this case the string rotates at the speed of light so it can be related to string theory with only one chiral sector.
Indeed the energy excitation $\epsilon$ contributes to the energy-momentum tensor $S_{ij}$ as
\beq
S^{\epsilon}_{ij} = C \left(
					  	\begin{array}{cc}
					    	1     & \pm 1 \\
					 		\pm 1 & 1
						\end{array}
					  \right) ,
\eeq{chiral}
for some constant $C$.
This has the same form as the energy-momentum tensor of string theory with $T_{++} =0$ (or $T_{--} = 0$), upon a coordinate transformation.
We do not have yet a full understanding of long string, but our results suggest that the study of their dynamics, 
even if limited to the classical level, can give a better understanding of gravity on AdS$_3$.

\section*{Acknowledgements}
J.K. is supported by an NYU JAGA fellowship. 
M.P. is supported in part by NSF grant PHY-1316452.


\end{document}